\documentclass[12pt]{article}

\usepackage{paper2e}
\usepackage{mydefs2e}

\newcommand{\LUV}{\Lambda_{\rm UV}}
\newcommand{\LC}{\Lambda_{\rm CFT}}
\newcommand{\MP}{M_{\rm Pl}}

\newcommand{\Lh}{\Lambda_{\rm hol}}

\newcommand{\vC}{v_{\rm CFT}}

\begin{document}
\def\Yfund{\drawsquare{7}{0.6}}

\newcommand{\ths}{\vartheta}

\newcommand{\N}{$\scr{N}=1$\xspace}

\begin{titlepage}
\preprint{UMD-PP-01-054 \\
JHU-TIPAC-2001-01}

\title{Supersymmetry Breaking\\\medskip
and Composite Extra Dimensions}

\author{Markus A. Luty}

\address{Department of Physics, University of Maryland\\
College Park, Maryland 20742, USA\\
{\tt mluty@physics.umd.edu}}

\author{Raman Sundrum}

\address{Department of Physics and Astronomy, Johns Hopkins University\\
Baltimore, Maryland 21218, USA\\
{\tt sundrum@pha.jhu.edu}}

\begin{abstract}
We study supergravity models in {\it four dimensions} where the
hidden sector is
superconformal and strongly-coupled over several decades of energy below the
Planck scale, before undergoing spontaneous breakdown of  scale
invariance and supersymmetry.
We show that large anomalous dimensions can suppress \Kahler contact terms
between the hidden and visible sectors, leading to models in which the hidden
sector is `sequestered' and anomaly-mediated supersymmetry breaking can
naturally dominate, thus solving the supersymmetric flavor problem.
We construct simple, explicit models of the hidden sector based on
supersymmetric QCD in the conformal window.
The present approach can be usefully interpreted as having an extra dimension
responsible for sequestering replaced by the many states of a
(spontaneously-broken) strongly-coupled superconformal hidden sector, as
dictated by the AdS/CFT correspondence.
\end{abstract}

\end{titlepage}

\section{Introduction}
The AdS/CFT correspondence \cite{adscft} asserts
that a gravity theory on 5D anti de Sitter space (AdS)
is `dual' to a 4D conformal field theory (CFT).
The duality takes the form of an equality of generating functionals
depending on some 4D fields $h_0$ that act as boundary
values for the gravity fields on AdS and source
terms for operators in the CFT:
\beq[AdSCFTeq]
\int\displaylimits_{\left. h \right|_{\rm bdy} = h_0} \!\!\!\!\!\!\!
d[h]\, e^{iS_{\rm grav}[h]} =
\left\langle e^{i \int h_0 \cdot {\cal O}} \right\rangle_{\rm CFT}.
\eeq
This correspondence remains at present an unproven conjecture, but in string
theory realizations it passes an impressive number of quantitative and
qualitative consistency checks, and has proved extremely fruitful in suggesting
new connections \cite{adscftrev}.

It has been argued in \Refs{adscftrs,apr,rz} 
that this duality can be
extended to
the equivalence between the 5D `brane world' scenario of Randall and Sundrum
(RS) \cite{rs1} and a conformal field theory perturbed by four-dimensional
gravity.
In this duality the `UV brane' in the RS model where gravity
is localized is mapped to a UV cutoff for the CFT and the redshifted
`IR brane' is mapped to spontaneous breaking of conformal invariance 
\cite{apr,rz}, which provides an IR cutoff of the CFT.
As stressed in \Ref{apr}, both sides of this perturbed duality
are macroscopically 4D theories with a discrete spectrum, and hence
the equivalence reduces simply to the statement that both theories
give identical predictions for all physical quantities such as the
S-matrix. This conjecture also passes several quantitative
checks \cite{gubser} and many qualitative ones \cite{apr,rz}.

In this paper, we will study 4D CFT's that can be viewed as being
dual to the RS model with supersymmetry (SUSY) \cite{susyrs,ls2,bagger}.
The phenomenological motivation for
this class of models is quite different from the original RS model, where the
redshift factor between the UV and IR branes was used to explain the hierarchy
between the Planck and weak scales.
In the SUSY RS model, it is SUSY that solves the hierarchy problem.
The motivation for the extra dimension comes from the SUSY flavor problem.
If we assume that there is no flavor symmetry at the Planck scale, then
the low-energy effective theory necessarily includes contact terms of the form
\beq[contact]
\De\scr{L}_{\rm eff} \sim \myint d^4\th\, \frac{1}{\MP^2} \Si^\dagger \Si
Q^\dagger Q,
\eeq
where $\Si$ is the hidden sector field that breaks SUSY and $Q$ is a
visible sector field.
Such terms cannot be forbidden by any symmetry, and give a contribution to the
squark masses of order $m_{3/2} \sim \avg{F_\Si}/\MP$ that has no reason to be
flavor diagonal.
In hidden sector models where soft SUSY breaking parameters are of order
$m_{3/2}$, it is therefore difficult to understand why the squark masses
are nearly flavor independent, as required by constraints on flavor-changing
neutral currents.
\Ref{rs0} showed that this problem can be solved if the visible and hidden
sectors are localized on different branes, separated in an extra dimension.
The spatial separation 
suppresses contact terms between
the hidden and visible sectors \cite{rs0,ls1}.
In \Ref{rs0}, this was referred to as `sequestering' the hidden sector.
If there are no massless fields in the bulk other than supergravity, SUSY
breaking  is communicated from the supergravity sector to the visible sector
by anomaly mediation \cite{rs0,glmr} (after radion
stabilization \cite{ls1}).%
\footnote{If the visible sector gauge fields propagate in the bulk,
this scenario leads to gaugino mediated SUSY breaking \cite{gmsb} or
radion mediated SUSY breaking \cite{rmsb}.
The dual
CFT description of these mechanism will be discussed elsewhere \cite{ls4}.}
Assuming that the visible sector contains only the MSSM and assuming no
other couplings between the visible and hidden sectors
implies that slepton mass-squared terms are negative,
but by relaxing these assumptions realistic and predictive
models have been constructed
that preserve the attractive features of the scenario \cite{amsbfix}.

The AdS/CFT correspondence extended to the SUSY RS set-up asserts that
anomaly-mediated models can be realized in a purely 4D theory.
The necessary ingredients in the 4D theory are determined simply by following
the correspondence.
The bulk supergravity modes in the 5D description are mapped to a
strongly-coupled superconformal field theory (SCFT) in the 4D theory.
Visible sector fields localized on the UV brane in the 5D description
are mapped
to elementary fields in the 4D theory that are coupled to the SCFT only through
Planck-suppressed operators.
Hidden sector fields localized on the IR brane in the 5D description are mapped
to composites of the SCFT that arise from the spontaneous breaking of conformal
invariance. Stabilization of the extra-dimensional radius is mapped to the
stabilization of the modulus in the CFT responsible for spontaneous
breakdown of
scale invariance.
Finally, the condition on the 5D theory that there are no light bulk
modes other
than supergravity responsible for transmitting supersymmetry breaking is mapped
to the condition that only irrelevant SCFT operators couple to the visible
fields.

We will be interested in strongly-coupled SCFT's with no expansion
parameters, such as large $N$ or large 't Hooft parameter.
AdS duals of such theories are not known, presumably because there
is no parametric separation between the string length, the AdS radius,
and the Planck length.
In order to achieve sequestering in such a theory by decoupling the
effects of massive bulk  states,
we require an extra dimension that extends over several
AdS radii.
At the level of effective field theory, the existence of
such strongly warped SUSY models was demonstrated in \Refs{susyrs}.
In \Ref{ls2} it was shown that radius stabilization and anomaly-mediated
SUSY breaking can be realized in this scenario.

The purpose of this paper is to show that sequestering and anomaly mediation 
are
indeed realized in a large class of 4D SUSY theories.
The theories can be explicitly constructed and understood
from a purely 4D perspective, and demonstrate that sequestering
can be realized without positing extra dimensions or branes.
However, we find the dual 5D description, where sequestering has a
simple geometrical origin, very illuminating.
We therefore refer to this class of models as `composite extra dimensions.'

\Ref{ns} considered SUSY models where the MSSM has superpotential
couplings to a strong SCFT and studied implications for flavor
and SUSY breaking.
\Refs{acg} constructed non-SUSY gauge theories whose low-energy
dynamics mimics that of a theory with an extra dimension.
We will briefly discuss the relation of these papers to our work
in the conclusions.

\section{CFT Suppression of Hidden-Visible Contact Terms}
The prototype of the kind of SCFT to which our results apply is the strongly
coupled fixed point of $SU(N)$ SUSY QCD with $F$ flavors
found by Seiberg \cite{Seibergdual}.
For $\sfrac{3}{2} N < F < 3N$, this theory is asymptotically free in
the UV but has a nontrivial conformal fixed point in the IR.
For $F \simeq 3N$ the IR fixed point is weakly coupled \cite{BanksZaks},
and for $F \simeq \sfrac{3}{2} N$ the IR fixed point has a weakly-coupled
dual description \cite{Seibergdual}, but in the middle of the range
the IR fixed point has no known weakly-coupled lagrangian description.
In what follows, we will write our results for the special case
$N = 2$, $F = 4$ for simplicity.
In that case there are 8 $SU(2)$ fundamentals $T^J$, $J = 1, \ldots, 8$.
We define $\LC$ to be the scale below which the theory is in the IR conformal
regime, and above which the theory rapidly runs to its asymptotically free
regime.

The crucial question is the size of scalar masses induced by flavor-violating
couplings of the form
\beq[CFTcontact]
\De\scr{L}_{\rm eff} =
\myint d^4\th\, \frac{c^j{}_k}{\MP^2} T_J^\dagger T^J
Q_j^\dagger Q^k,
\eeq
where $T$ is a hidden sector SUSY QCD field.
Note that we have assumed that the coupling is diagonal in hidden flavor,
which can be made natural by imposing (discrete and/or gauged) flavor
symmetries
on the hidden sector.
In order for anomaly mediation to dominate, we require that this term
contribute  visible scalar masses
$\De m_{\tilde{Q}}^2 \lsim 10^{-7} V_{\rm hid} / \MP^2$ (see Section 3),
where $V_{\rm hid}$ is the SUSY breaking vacuum energy.
We will discuss the mechanism of this SUSY breaking in the hidden sector
in the next section.
The suppression factor of $10^{-7}$ must arise from the nontrivial
CFT scaling of the operator \Eq{CFTcontact}.

This term can be viewed as a
correction to the kinetic term for the $T$ fields in the UV lagrangian
\beq[badop]
\scr{L}_{\rm UV} = \myint d^4\th\, Z_0
T_J^\dagger T^J
+ \cdots,
\qquad
Z_0 = 1 + \frac{c^j{}_k}{\MP^2} Q^\dagger_j Q^k.
\eeq
This contributes to a perturbation of the physical gauge coupling $g^2$,
given by \cite{NSVZetc}
\beq[gphys]
\frac{1}{g^2} = \frac{1}{g_{\rm hol}^2}
- \frac{N}{8\pi^2} \ln g^2
- \frac{F}{8\pi^2} \ln Z
+ \hbox{\rm constant} + \scr{O}(g^2),
\eeq
where $1/g^2_{\rm hol}$ is the holomorphic gauge coupling that appears as
the coefficient of the gauge kinetic term in the lagrangian.
Because \Eq{badop} is a perturbation to the UV gauge coupling,
it is necessarily irrelevant near the fixed point.
This is simply because the
theory near a fixed point must be  insensitive to UV couplings in order
to be IR attractive, as Seiberg argued is the case in SUSY QCD.


To make this quantitative, let us consider how the operator 
\Eq{badop} runs down
to the IR in two stages.
First, the running down to the scale $\LC$ where the theory becomes strong is
the standard logarithmic running in the far UV, together with an order unity
strong-interaction renormalization near $\LC$.
$\LC$ is defined such that $g(\LC)$ is a {\it fixed number} close 
enough  to the
fixed point coupling $g_*$ that below this scale  we can expand about the
 fixed
point:
\beq
\be = \be'_* \cdot (g^2 - g_*^2) + \cdots,
\qquad
\ga = \ga_* + \ga'_* \cdot (g^2 - g_*^2) + \cdots,
\eeq
where $\be \equiv d g^2 / d\ln\mu$, $\ga \equiv d \ln Z / d\ln\mu$.
The anomalous dimension at the fixed point is determined by the (non-anomalous)
$U(1)_R$ symmetry to be
\begin{equation}
\ga_* = \sfrac{1}{2}.
\end{equation}
  Integrating these renormalization group equations from $\LC$ down
to $\mu$ then gives
\beq\!\!\!\!\!\!\!\!\!\!\!\!\!
  Z(\mu) = Z(\LC) \left( \frac{\mu}{\LC} \right)^{\gamma_*}
  \left\{
  1 + \frac{\ga'_*}{\be'_*} \left[ g^2(\LC) - g_*^2 \right]
  \left[\left( \frac{\mu}{\LC} \right)^{\be'_*} - 1 \right] 
  + \cdots \right\}.
\eeq
  We can rewrite this using \Eq{gphys} evaluated at $\LC$ and the fact that
  $g_{\rm hol}$ has exact one-loop running,
\beq[Zmu]\!\!\!\!\!\!\!\!\!\!\!\!\!
  Z(\mu) = \hbox{\rm const} \times
\left( \frac{\mu}{|\Lh|} \right)^{\gamma_*} \left\{
1 + \frac{\ga'_*}{\be'_*} \left[ g^2(\LC) - g_*^2 \right]
 \left[\left( \frac{\mu}{\LC} \right)^{\be'_*} - 1 \right] 
  + \cdots \right\},
\eeq
where
\beq
  \Lh \equiv \mu e^{-4\pi^2 / g^2_{\rm hol}(\mu)}
\eeq
  is the holomorphic one-loop strong-interaction scale.

\Eq{Zmu} is useful because it shows that the leading dependence on
$Z_0$ (contained in $Z(\LC)$) has disappeared in the IR below $\LC$.
The subleading dependence on $Z_0$ is implicit in the
dependence on $\LC$, which is suppressed by the power $\be'_* > 0$.
(Note that $\LC$ depends on $Z_0$ through its definition,
$g^2(\LC) = \hbox{\rm const}$, and also \Eq{gphys}.)
The fixed point behavior is cut off at the scale $\mu = \vC$ where
the conformal symmetry is spontaneously broken, so we obtain the
required suppression for
$(\vC / \LC)^{\be'_*} \lsim 10^{-7}$.

It is crucial that the perturbation \Eq{badop} is a singlet under the
hidden sector flavor symmetries.
For a non-singlet operator of the form
\beq[badbadop]
\De\scr{L} = \myint d^4\th 
\frac{c^{j J}{}_{k K}}{\MP^2}  T^\dagger_J T^K 
Q^\dagger_j Q^k,
\eeq
with
$c^{j J}{}_{k J} = 0$,
the contribution to the visible sector scalar masses is
\beq
(\De m^2_{\tilde Q})^j{}_k
= -\frac{c^{j J}{}_{k K}}{\MP^2} \left\langle
\myint d^4\th\, T^\dagger_J T^K \right\rangle
\eeq
This is a matrix element of a conserved
current supermultiplet with vanishing anomalous dimension,
so there is no CFT suppression of operators of the form \Eq{badbadop}.
A model-building requirement is therefore to insist on enough symmetry in the
hidden dynamics to prohibit such operators.

In the SUSY limit the theory above has a moduli space of vacua, and
away from the origin of moduli space the conformal symmetry is
spontaneously broken.
The light fields below the scale $\vC$ where the conformal symmetry
is broken are the moduli, which can be thought of as composites of the CFT.
We now consider the effective field theory for these moduli.
The moduli space can be parameterized by the gauge-invariant
holomorphic operators \cite{LT}.
In the $SU(2)$ gauge theory we are considering,
the moduli space is parameterized completely by the `meson'
invariants
\beq[MJK]
M^{JK} = T^J T^K = -M^{KJ}.
\eeq

   From \Eq{Zmu} we see that the dependence on $\Lh$ can be eliminated by
working in terms of the renormalized fields
\beq[Trenorm]
T'^J = \frac{T^J}{(\Lh)^{1/4}}.
\eeq
Since $\Lambda_{\rm hol}$ parametrizes the only explicit scale in the hidden
dynamics, and since the new fields eliminate dependence on this scale near
the IR fixed point,
the leading low-energy interactions must be given by
\beq[Leffmod]
\scr{L}_{\rm eff} = \myint d^4\th\, f(M', M'^\dagger)
+ \scr{O}(\partial^4)
+ \scr{O}(\LC^{-\be_*'}),
\eeq
where $M' = T' T'$ and
$f$ is a homogeneous function with its degree determined by dimensional
analysis:
\beq
f(\al M', \al M'^\dagger) = \al^{4/3} f(M', M'^\dagger).
\eeq
The new fields are very  convenient later  because they have the same
canonical dimension as their scaling dimension in superpotential terms,
allowing us to simultaneously non-linearly realize the asymptotic
(canonical) scale invariance in the UV and the non-trivial asymptotic
scale invariance in the IR.


There is another way to derive the absence of $Z_0$ dependence in the
leading terms of the low-energy theory, \Eq{Leffmod},
   that may be illuminating.
We can regard  $Z_0$ as a background gauge connection
for an anomalous $U(1)_A$ symmetry \cite{connection}.
Because of the anomaly, $\Lh$ is charged under this symmetry with a
charge such  that the renormalized fields are uncharged under
$U(1)_A$.
The leading terms in the low-energy effective theory are therefore
independent of the $U(1)_A$ gauge connection.
By contrast, perturbations of the form \Eq{badbadop} can be
regarded as background gauge connections for non-anomalous hidden
flavor symmetries. Therefore $\Lambda_{\rm hol}$ is uncharged under these
gauge connections and cannot cancel their effects in the low-energy theory,
\Eq{Leffmod}.

\section{A Realistic Model}
We now show how to construct a
realistic 4D model in which SUSY breaking is communicated
by anomaly mediation, with the suppression of contact terms explained
by the mechanism described above.
Our aim is to construct a model which illustrates the issues in
constructing a realistic model, and separates these issues as clearly
as possible.
The model we discuss 
contains several explicit small superpotential couplings whose origin
is not explained.
We believe that completely natural models without fundamental small
parameters are possible, but the leave their construction for future
work.

The hidden sector will be taken to be $SU(2)$ gauge theory with 8 fundamentals
$T^J$ ($J = 1, \ldots, 8)$, as discussed in the last section.
%
The classical moduli space of this theory can be parameterized by the
`meson' operators $M^{JK}$ (see \Eq{MJK}).
$M$ is an antisymmetric matrix with rank 2, which can be
conveniently parameterized by
\beq[moduli1]
M = \bordermatrix{
      & 2     & 6 \cr
2 &  X \ep & -Y^T \cr
6 & Y     & \scr{O}(Y^2 / X) \cr},
\qquad \ep = \pmatrix{0 & 1 \cr -1 & 0 \cr},
\eeq
where $X$ and $Y$ are unconstrained.
Note that this has 13 complex degrees of freedom, as required
given the sixteen quark fields and the three D-flatness conditions.
We will expand about the vacuum
\beq
\avg{M} =
\bordermatrix{
      & 2       & 6 \cr
2 & \avg{X} \ep & 0 \cr
6 & 0       & 0 \cr}.
\eeq
It will be convenient to further parameterize
\beq[moduli2]
Y = \bordermatrix{
      & 2 \cr
2 &  \Si\ep + \Pi \cr
4 & \Pi' \cr},
\qquad
\tr(\ep \Pi) = 0.
\eeq
Upon adding superpotential terms
we will see that SUSY is broken by $\avg{F_\Si} \ne 0$.

The low-energy effective theory below the scale determined by the
VEV $\avg{M}$ was described in the previous section.
The contribution to the effective \Kahler potential is 
a homogeneous function of the meson fields of
degree $\sfrac{4}{3}$, which must also be $SU(8)$ invariant.
We will always work in terms of the `renormalized' primed fields of the
previous section, dropping the primes.
The VEV breaks the $SU(8)$ global symmetry to $SU(2) \times SU(6)$,
so expanding in powers of $Y$ gives
\beq[Keff2]
K_{\rm eff} = a_0 (X^\dagger X)^{2/3} \left[
1 + a_1 \frac{\tr(Y^\dagger Y)}{X^\dagger X} + \scr{O}(|Y|^4/|X|^2)
\right],
\eeq
where $a_{0, 1}$ are unknown strong interaction parameters.
It is convenient to work in terms of redefined fields
\beq[canmoduli]
\hat{X} = a_0^{1/2} X^{3/2},
\qquad
\hat{Y} =  a_0^{3/4} a_1^{1/2} \frac{Y}{\hat{X}^{1/2}},
\eeq
which have a canonical \Kahler potential:
\beq
K_{\rm eff} = \hat{X}^\dagger \hat{X} + \hat{Y}^\dagger \hat{Y}
+ \scr{O}(|\hat{Y}|^4 / |\hat{X}|^2).
\eeq

As described so far, the model has unbroken SUSY and a moduli space of vacua.
We now add superpotential terms that stabilize the moduli and break
SUSY.
We will take our model, and the superpotential in particular,
    to respect  an $SU(2)$ subgroup of the
global $SU(8)$ symmetry. (This flavor $SU(2)$ symmetry can be weakly gauged
but we will not consider this here.)
For convenience we give names to the four $SU(2)$ doublets as follows:
\beq
P^{1,2} = T^{1,2},
\quad
\bar{P}^{1,2} = T^{3,4},
\quad
N^{1,2} = T^{5,6},
\quad
\bar{N}^{1,2} = T^{7,8},
\eeq
so that
\beq
X = P^j P_j,
\qquad
\Si = P^j \bar{P}_j,
\eeq
where $j, k = 1,2$ are global $SU(2)$ indices,
and we have defined $P_j \equiv \ep_{jk} P^k$, {\it etc\/}.
In addition to the global $SU(2)$ we impose the following discrete
symmetries:
\beq
(i)\ \ & P \leftrightarrow \bar{P},
\quad
N \leftrightarrow \bar{N},
\\
(ii)\ \ & P \leftrightarrow N,
\quad
\bar{P} \leftrightarrow \bar{N},
\\
(iii) \ \ & P \mapsto iP,
\quad
\bar{P} \mapsto -i \bar{P},
\quad
N \mapsto i N,
\quad
\bar{N} \mapsto -i\bar{N},
\\
(iv) \ \ & P \mapsto iP,
\quad
\bar{P} \mapsto -i \bar{P},
\quad
N \mapsto -i N,
\quad
\bar{N} \mapsto i\bar{N}.
\eeq
These symmetries ensure that the only allowed term of the form
$T^\dagger T$ is (accidentally) $SU(8)$ invariant.
Therefore the only \Kahler term of the form $T^\dagger T Q^\dagger Q
/ \MP^2$ is
a singlet of the CFT flavor symmetries, and is suppressed by the
renormalization group arguments
of the previous section.

These symmetries allow us to add the following terms to the superpotential:
\beq
W = W_{\rm stab} + W_{\rm mass} + W_{\rm Polonyi},
\eeq
where
\beq
& \bal
W_{\rm stab} &= \sfrac{1}{2} \la_1 \left[ (P^j P_j)^2 + (\bar{P}^j \bar{P}_j)^2
+ (N^j N_j)^2 + (\bar{N}^j \bar{N}_j)^2 \right]
\\
&\qquad
+\, \sfrac{1}{4} \la_2 \left[ (P^j P_j)^4 + (\bar{P}^j \bar{P}_j)^4
+ (N^j N_j)^4 + (\bar{N}^j \bar{N}_j)^4 \right],
\eal
\\
& \bal
W_{\rm mass} &= c_1 \sum_{a = 1}^{3}
\left[ (P^j \si^a{}_j{}^k \bar{P}_k)^2 + (N^j \si^a{}_j{}^k \bar{N}_k)^2
\right]
\\
& \qquad
+\, c_2 \left[
(P^j N^k) (P_j N_k)
+ (\bar{P}^j \bar{N}^k) (\bar{P}_j \bar{N}_k) \right.
\\
& \qquad\qquad\quad
\left.
+ (\bar{P}^j N^k) (\bar{P}_j N_k)
+ (P^j \bar{N}^k) (P_j \bar{N}_k)
\right],
\eal
\\
& W_{\rm Polonyi} = \ka \left[
P^j \bar{P}_j + N^j \bar{N}_j \right],
\eeq
and where $\si^{1,2,3}$ are the Pauli matrices for flavor $SU(2)$.
Gauge-singlet operators are enclosed by parentheses.
At the scale $\LC$ the $SU(2)$ gauge interactions become strong
and the theory flows to a nontrivial conformal fixed point.
At this point the scaling of the operators is controlled by the nontrivial
fixed point.
We will assume that
\begin{equation}
\LC \sim \LUV \sim 4\pi\MP,
\end{equation}
where $\LUV$ is the scale where
4D quantum gravity becomes strong.
We will work in units where $\MP = 1$.

Below the scale $\avg{\hat{X}}$ the conformal symmetry is spontaneously
broken and the effective degrees of freedom of the CFT are the moduli.
Writing the superpotential in terms of the moduli fields defined in
\Eqs{moduli1}, \eq{moduli2}, and \eq{canmoduli} we have
\beq
\eql{Wstab}
& \bal
W_{\rm stab} &= \sfrac 12 \la_1 \left[ \hat{X}^3
+ \scr{O}(\hat{\Si}^4 / \hat{X}) + \scr{O}(\hat{\Pi}^4 / \hat{X}) \right]
\\
&\qquad
+\, \sfrac 14 \la_2 \left[ \hat{X}^6
+ \scr{O}(\hat{\Si}^8 / \hat{X}^2) + \scr{O}(\hat{\Pi}^6 / \hat{X}^2) \right],
\eal
\\
\eql{WPoleff}
& W_{\rm Polonyi} = \ka \hat{X}^{1/2} \hat{\Si},
\eeq
while $W_{\rm mass}$ is a sum of mass terms for every component of $\Pi$ and
$\Pi'$.
In the above, we have absorbed the unknown strong interaction coefficients
of the \Kahler potential into redefinitions of the superpotential
couplings (see \Eqs{Keff2} and \eq{canmoduli}).

The CFT running of the
superpotential perturbations is automatically taken into account
by our working in terms of the primed fields defined in the previous
section.
Our discussion assumes that the superpotential terms can be treated
as linear perturbations to the CFT in the energy range from $\LUV$
to $\avg{\hat X}$.
The couplings $\la_1$, $c_1$, and $c_2$ are marginal and have dimensionless
coefficients small compared to 1.
There are nonlinear corrections suppressed by higher powers of these
couplings, but these are negligible logarithmic corrections similar to
those found in weak-coupling perturbation theory.
The coupling $\la_2$ is irrelevant, and can therefore also be treated
as a perturbation.

We now determine the VEV's.
The stabilization term \Eq{Wstab} gives rise to a local SUSY preserving
minimum with
\beq[X]
\avg{\hat{X}}^3 = -\frac{\la_1}{\la_2}.
\eeq
(We will consider the effects of SUSY breaking on $\hat{X}$ below.)
This stabilizes the modulus and determines the scale of spontaneous
breaking of the conformal symmetry.
The mass of the $X$ field is of order
\beq
m_X \sim \la_1 \avg{\hat{X}}.
\eeq
Since $\la_2 \lsim 1$ and we want $\avg{\hat{X}} \ll 1$ (in Planck units)
we must have $\la_1 \ll 1$, and hence $m_X \ll \avg{\hat{X}}$.
We will take $\la_2 \sim 1$ in what follows.%
\footnote{This is in fact conservative, since \naive dimensional
analysis \cite{nda} allows a larger coefficient.}

Below the scale $m_X$ we integrate out $X$ and consider the effective
lagrangian of the remaining light degrees of freedom.
We chose the couplings in $W_{\rm mass}$ so that all of the modes
parameterized by $\Pi$ and $\Pi'$ get masses $m_\Pi \lsim m_X$.
The only remaining degree of freedom below the scale $m_\Pi$ is then
$\Si$.
The effective superpotential is then
\beq
W_{\rm eff} = \ka \avg{\hat{X}}^{1/2} \hat{\Si}
+ \scr{O}(\ka^2 \hat{\Si^2} / \avg{\hat{X}}^5),
\eeq
where the terms higher order in $\hat\Si$ come from the $\hat{X}$
dependence in \Eq{WPoleff}.
The effective \Kahler potential is
\beq
K_{\rm eff} = \hat{\Si}^\dagger \hat{\Si}
+ c \frac{(\hat{\Si}^\dagger \hat{\Si})^2}{|\avg{\hat{X}}|^2} + \cdots,
\eeq
where $c$ is an order one unknown strong interaction coefficient.
If  $c < 0$, this theory has a (local) SUSY breaking minimum with
\beq
\avg{F_{\hat\Si}} \sim \ka \avg{\hat{X}}^{1/2} 
\left[ 1 +
{\cal O}(\kappa^2 / \hat{X}^9) \right],
\qquad
\avg{\hat\Si} = {\cal O}(\kappa / \hat{X}^{7/2}),
\eeq
and $\Si$ gets a mass
\beq
m_\Si \sim \frac{\avg{F_{\hat\Si}}}{\avg{\hat{X}}}.
\eeq
Here we will simply make the dynamical assumption that $c < 0$.
The condition that the higher order terms make only a small fractional
correction to the SUSY breaking order parameter $F_{\hat\Sigma}$ is
\beq[Fcond]
\frac{\avg{F_{\hat\Sigma}}^2 M_{\rm Pl}^6}{\avg{\hat{X}}^{10}} \lsim 1.
\eeq
This discussion assumes that $\hat{X}$ is sufficiently heavy that we can
integrate it out for purposes of SUSY breaking.
We also assumed that SUSY breaking does not significantly shift the
$\avg{\hat{X}}$ away from its SUSY value.
It is easily checked that both of these constraints are equivalent
to \Eq{Fcond}.

We are now ready to check the numbers.
The most stringent constraints on flavor-changing neutral currents
arise from $K^0$--$\bar{K}^0$ mixing \cite{flavor}:
\beq
\frac{m^2_{\tilde{d}\tilde{s}}}{m^2_{\tilde{s}}}
\lsim (6 \times 10^{-3}) \left( \frac{m_{\tilde{s}}}{1\TeV} \right),
\qquad
\Im \left(\frac{m^2_{\tilde{d}\tilde{s}}}{m^2_{\tilde{s}}} \right)
\lsim (4 \times 10^{-4}) \left( \frac{m_{\tilde{s}}}{1\TeV} \right).
\eeq
We assume squark masses of order $1\TeV$ in the following.
Anomaly mediation gives a flavor-independent mass to squarks of order
\beq
m_{\tilde{q}} \sim 2 \times 10^{-2} F_\phi,
\eeq
where $F_\phi \sim m_{3/2} \sim \avg{F_{\hat\Si}} / \MP$.
This fixes
\beq
\avg{F_{\hat\Si}} \sim 8 \times 10^{22}\GeV^2.
\eeq
The required suppression of FCNC's is obtained provided
\beq[FCNCAMSBconst]
\left( \frac{\avg{\hat{X}}}{\LUV} \right)^{\be'_*} \lsim
2 \times 10^{-7}
\eeq
using the stronger CP-violating constraint.
The constraint \Eq{Fcond} then gives
\beq
\avg{\hat X} \gsim 4 \times 10^{15} \GeV.
\eeq
The constraint \Eq{FCNCAMSBconst} is satisfied provided
\beq[beta]
\be'_* \gsim 1.7.
\eeq
As discussed above, $\be'_*$ is a non-perturbatively determined exponent
which we cannot calculate.
Naive dimensional analysis \cite{nda} tells us that $\be'_* \sim 1$.
Extrapolations using perturbation theory valid for the
    Banks-Zaks fixed points \cite{BanksZaksextrap}, 
$1 - F/(3N) \ll 1$, suggests
that $\be'_* \simeq 1$ at the self-dual point, $F = 2N$.
In the absence of more rigorous information, we believe that values
such as this are very reasonable.
In fact we are able to construct models that allow smaller values of $\be'_*$
than \Eq{beta} by using stabilizing superpotentials with smaller powers of $X$.
In the present model, such powers are forbidden by discrete symmetries, but we
can add more fields and couplings that spontaneously break these symmetries and
generate lower powers of $X$ below the breaking scale.
The analysis of such models is slightly more involved than our
present model and
will not be detailed here.

Another dynamical assumption required in this model is that the uncalculable
strong \Kahler corrections have the right sign ($c < 0$)
to give a local SUSY breaking vacuum at $\avg{\hat\Si} = 0$.
This dynamical assumption can avoided by replacing $W_{\rm Polonyi}$ by
an O'Raifeartaigh sector with additional singlet fields.
Basically, the additional fields in the O'Raifeartaigh sector give
larger calculable \Kahler corrections than the
uncalculable \Kahler corrections
if these singlets are sufficiently light.
If some of the additional singlet fields are elementary, one must
ensure that they do not get substantial $F$ terms, since (standard
model flavor violating) contact interactions
between these fields and the visible fields are not suppressed.
We have constructed explicit models of this type.

So far we have considered the hidden dynamics in flat spacetime, showing how
to stabilize the moduli and break SUSY. Because the energy scales and
VEVs in the hidden sector are much smaller than the Planck scale, the main
effect of coupling the hidden sector to supergravity is that the SUSY
breaking contribution to the cosmological constant can be cancelled in the
usual way. Supergravity has a very small effect on the hidden dynamics
and vacuum stabilization. The main effect of coupling supergravity to the
visible sector is that SUSY breaking is communicated to the visible sector
by the mechanism of anomaly-mediation.

It is straightforward to adapt proposals in the literature \cite{amsbfix}
for solving
the tachyonic slepton problem and $\mu$-problem of anomaly mediation
to the present framework.


Note that the stabilizing superpotential has another
supersymmetric solution, $X = 0$. At this point on the moduli space
the theory remains superconformal and supersymmetric, and therefore
has lower energy than the local minimum, \Eq{X}. In other words
our supersymmetry-breaking vacuum is only meta-stable. However we have
checked that tunneling to the true supersymmetry-preserving vacuum is highly
suppressed over cosmological time scales, just as in the SUSY breaking
scenario of \Ref{cosmo}.

\section{Conclusions}
The main result of this paper is that it is possible to construct
4D SUSY field theories that realize the sequestering of the hidden
sector.
The original sequestering mechanism of \Ref{rs0} had its origin in
the spatial separation of the visible and hidden sectors in an
extra dimension.
In the 4-dimensional models considered here the role of the extra dimension
is played by the many states of a superconformal field theory,
as dictated by the AdS/CFT correspondence.
Using these ideas we have constructed an explicit realistic 4D model in
which anomaly mediation dominates in the visible sector.

The higher-dimensional realization of sequestering is geometric
and highly intuitive.
However, the local higher-dimensional ($\scr{N} = 2$) SUSY is a
significant technical complication that makes the construction of
explicit models difficult.
In the 4D models considered here the extra supersymmetry is
implicit in the enhanced superconformal symmetry  of the fixed point,
and we only need to keep track of $\scr{N} = 1$ SUSY for
model-building.

There are many interesting further directions to pursue.
In future work \cite{ls4}, we intend to extend the ideas of this paper
to study 4D realizations
of gaugino mediation \cite{gmsb} and radion mediation \cite{rmsb},
where the hidden sector has a superconformal regime (dual to having the
hidden sector on the IR brane in a SUSY RS set-up).
We also wish to consider the important question of constructing
fully natural models with a dynamical origin for scale hierarchies.

We end with some comments on related work that has appeared recently.
\Ref{ns} studied strong SUSY CFT's applied to the flavor problem and SUSY
breaking.
Although the relation to the AdS/CFT correspondence was not discussed in this
paper, for purposes of SUSY breaking the models of \Ref{ns} can be
viewed as the
CFT dual of gaugino mediation \cite{gmsb}, with the hidden sector localized on
the UV brane.
There are some important differences between this work and that
of the present paper, beyond the obvious difference in the SUSY-breaking 
mediation mechanism. 
In \Ref{ns} the conformal symmetry is broken by relevant operators, and
suppressing all soft terms requires flavor symmetries in the standard model to
be completely broken.
While the scenario of \Ref{ns} implements a specific proposal for understanding
the structure of Yukawa couplings as well as giving a solution to the SUSY
flavor problem, our present work is aimed only at the SUSY flavor problem.
On the other hand realistic model-building  appears to be simpler in the present
approach where the hidden sector originates from the CFT (dual to having the
SUSY breaking on the IR brane).

\Refs{acg} gave a very simple and 
explicit construction of gauge theories whose low-energy dynamics
mimics that of a flat extra dimension without gravity. 
In this approach the many states of the extra dimension arise from
having many four-dimensional gauge sectors, while in our approach
they arise from the excited states of a simple
CFT.
In the framework of \Refs{acg} sequestering is difficult to realize because it is
not clear how to maintain locality in the the extra-dimensional
interpretation in the presence of gravity.

\section*{Acknowledgements}
M.A.L. was supported by NSF grant PHY-98-02551.

\newpage


\begin{thebibliography}{99}
\frenchspacing


\bibitem{adscft} J.~Maldacena,
{\it Adv.\ Theor.\ Math.\ Phys.}\ {\bf 2} (1998) 231, {\tt hep-th/9711200};
S.S.~Gubser, I.R.~Klebanov, A.M.~Polyakov,
{\it Phys.\ Lett.}\ {\bf B428} (1998) 105, {\tt hep-th/9802109};
E.~Witten, {\it Adv.\ Theor.\ Math.\ Phys.}\ {\bf 2}
(1998) 253, {\tt hep-th/9802150}.

\bibitem{adscftrev}
For a review see O.\ Aharony, S.S.\ Gubser, J.\ Maldacena, H.\ Ooguri, Y.\ Oz,
{\em Phys.\ Rept.}\ {\bf 323} (2000) 183,
{\tt hep-th/9905111}.




\bibitem{adscftrs}
H. Verlinde, {\em  Nucl. Phys.} {\bf  B580} (2000) 264,
{\tt hep-th/9906182};
J. Maldacena, unpublished remarks;
E. Witten, ITP Santa Barbara conference
`New Dimensions in Field Theory and String Theory,',
{\tt http://www.itp.ucsb.edu/online/susy
c99/discussion};
E.~Verlinde, H.~Verlinde, {\em  JHEP} {\bf 0005} (2000) 034, 
{\tt hep-th/9912018}. 

\bibitem{apr}
N. Arkani-Hamed, M. Porrati, L. Randall, {\tt hep-th/0012148}.

\bibitem{rz}
R. Rattazzi, A. Zaffaroni,
{\em JHEP} {\bf 0104} (2001) 021,
{\tt hep-th/0012248}.


\bibitem{rs1}
L.~Randall, R.~Sundrum,
{\em Phys.\ Rev.\ Lett.}\ {\bf 83} (1999) 3370, {\tt hep-ph/9905221};
{\em Phys.\ Rev.\ Lett.}\ {\bf 83} (1999) 4690, {\tt hep-th/9906064}.


\bibitem{gubser}
S. S. Gubser, {\em Phys.\ Rev.}\ {\bf D63} (2001) 084017,
{\tt hep-th/9912001};
M. Perez-Victoria, {\tt hep-th/0105048}.



\bibitem{susyrs}
M. Cvetic, H. Lu, C.N. Pope, {\tt hep-th/0002054};
R. Altendorfer, J. Bagger, D. Nemeschansky, {\tt hep-th/0003117};
T. Gherghetta, A. Pomarol, {\tt hep-ph/0003129};
N. Alonso-Alberca, P. Meessen, T. Ortin, {\em Phys. Lett.}
{\bf B482} (2000) 400, {\tt hep-th/0003248};
A. Falkowski, Z. Lalak, S. Pokorski, {\tt hep-th/0004093};
E. Bergshoeff, R. Kallosh, A. Van Proeyen,
{\em JHEP} 0010 (2000) 033, {\tt hep-th/0007044};
M. Zucker, {\tt hep-th/0009083};
H. Nishino, S. Rajpoot, {\tt hep-th/0011066}.


\bibitem{ls2}
M.A. Luty, R. Sundrum, {\tt hep-th/0012158}.


\bibitem{bagger} 
J.~Bagger, D.~Nemeschansky, R.-J.~Zhang, {\tt hep-th/0012163}. 


\bibitem{rs0}
L. Randall, R. Sundrum, {\em Nucl.\ Phys.}\ {\bf B557} (1999) 79,
{\tt hep-th/9810155}.

\bibitem{ls1}
M.A.~Luty, R.~Sundrum,
{\em Phys.\ Rev.}\ {\bf D62} (2000) 035008,
{\tt hep-th/9910202}.

\bibitem{glmr}
G.F. Giudice, M.A. Luty, H. Murayama, R. Rattazzi,
{\em JHEP} {\bf 9812} (1998) 027, {\tt hep-ph/9810442}.






\bibitem{gmsb}
D.E. Kaplan, G.D. Kribs, M. Schmaltz, {\em Phys.\ Rev.}\ {\bf D62}
(2000) 035010, {\tt hep-ph/9911293};
Z. Chacko, M.A. Luty, A.E. Nelson, E. Pont\'{o}n,
{\em JHEP} {\bf 0001} (2000) 003, {\tt hep-ph/9911323}.

\bibitem{rmsb}
Z. Chacko, M.A. Luty, {\tt hep-ph/0008103}.

\bibitem{ls4}
M.A. Luty, R. Sundrum, in preparation.






\bibitem{amsbfix}
A.~Pomarol and R.~Rattazzi,
{\em JHEP} {\bf 9905} (1999) 013, {\tt hep-ph/9903448};
Z. Chacko, M.A. Luty, I. Maksymyk, E. Pont\`on,
{\em JHEP} {\bf 0004} (2000) 001,
{\tt hep-ph/9905390};
E. Katz, Y. Shadmi, Y. Shirman {\em JHEP} {\bf 9908} (1999) 015,
{\tt hep-ph/9906296};
K. I. Izawa, Y. Nomura, T. Yanagida, {\em Prog. Theor. Phys.} {\bf 102} (1999)
1181, {\tt hep-ph/9908240};
M. Carena, K. Huitu, T. Kobayashi, {\em Nucl. Phys.} {\bf B592} (2000) 164,
{\tt hep-ph/0003187};
B.C.~Allanach, A.~Dedes,
 {\em JHEP} {\bf 0006} (2000) 017, {\tt hep-ph/0003222};
I. Jack, D.R.T. Jones, {\em Phys. Lett.} {\bf B491} (2000) 151, {\tt
hep-ph/0006116};
D.E. Kaplan, G.D. Kribs, {\em JHEP} 0009 (2000) 048, {\tt hep-ph/0009195}.
N.~Arkani-Hamed, D.~E.~Kaplan, H.~Murayama, Y.~Nomura,
{\em JHEP} {\bf 0102} (2001) 041,
{\tt hep-ph/0012103}.







\bibitem{ns}
A.E.~Nelson, M.J.~Strassler, {\tt hep-ph/0104051};
{\em JHEP} {\bf 0009} (2000) 030,
{\tt hep-ph/0006251}.

\bibitem{acg}
N.~Arkani-Hamed, A.G.~Cohen, H.~Georgi,
{\tt hep-th/0104005};
C.T. Hill, S. Pokorski, J. Wang, {\tt hep-th/0104035}.






\bibitem{Seibergdual}
N. Seiberg, {\em Nucl.\ Phys.}\ {\bf B435} (1995) 129,
{\tt hep-th/9411149}.

\bibitem{BanksZaks}
T.~Banks, A.~Zaks,
{\em Nucl.\ Phys.}\ {\bf B196} (1982) 189.






\bibitem{NSVZetc}
V.A. Novikov, M.A. Shifman, A.I. Vainshtein, V.I. Zakharov,
{\em Nucl.\ Phys.}\ {\bf B229} (1983) 381;
M.A. Shifman, A.I. Vainshtein, {\em Nucl.\ Phys.}\ {\bf B277} (1986) 456;
{\em Nucl.\ Phys.}\ {\bf B359} (1991) 571;
N. Arkani-Hamed, H. Murayama, {\em JHEP} {\bf 0006} (2000) 030,
{\tt hep-th/9707133};
N. Arkani-Hamed, G. Giudice, M.A. Luty, R. Rattazzi,
{\em Phys.\ Rev.}\ {\bf D58} (1998) 115005,
{\tt hep-ph/9803290}.





\bibitem{LT}
M.A. Luty, W. Taylor IV, {\em Phys.\ Rev.}\ {\bf D53} (1996) 3399,
{\tt hep-th/9506098}.

\bibitem{connection}  N.~Arkani-Hamed, R.~Rattazzi, {\em Phys.\ Lett.} 
{\bf B454} (1999) 290, 
{\tt hep-th/9804068}; M.A.~Luty, R.~Rattazzi, {\em JHEP} {\bf 9911}
(1999) 001,  {\tt  hep-th/9908085}.  




\bibitem{BanksZaksextrap}
E.~Gardi, G.~Grunberg, {\em JHEP} {\bf 9903} (1999) 024,
{\tt hep-th/9810192}.

\bibitem{nda} A.~Manohar, H.~Georgi, {\em Nucl.\ Phys.}\ {\bf B234}
(1984) 189; H.~Georgi, L.~Randall, {\em Nucl.\ Phys.}\ {\bf B276}
(1986) 241.

\bibitem{flavor} F.~Gabbiani, E.~Gabrielli, A.~Masiero, 
L.~Silvestrini, {\em Nucl.\ Phys.}\ {\bf B477}
(1996) 321, {\tt hep-ph/9604387}.

\bibitem{cosmo} S.~Dimopoulos, G.~Dvali, R.~Rattazzi, G.F.~Giudice,
{\em Nucl.\ Phys.}\ {\bf B510}
(1998) 12, {\tt  hep-ph/9705307}. 




\end{thebibliography}
\end{document}